\documentclass[aps,prb,showpacs,twocolumn,amssymb,floatfix]{revtex4}

\usepackage{amsmath}
\usepackage{graphicx}
\usepackage{longtable}

\begin{document}
  \bibliographystyle{prsty}
  \title{Adsorption geometry of Cu(111)-Cs studied by scanning tunneling microscopy}
  \author{Th.\ von Hofe}
  \author{J.\ Kr\"oger}\email{kroeger@physik.uni-kiel.de}
  \author{R. Berndt}
  \affiliation{Institut f\"ur Experimentelle und Angewandte Physik, Christian-Albrechts-Universit\"at zu Kiel, D-24098 Kiel, Germany}

  \begin{abstract}
    Using scanning tunneling microscopy at low temperatures we investigated
    cesium adsorbed on Cu(111). At low coverages we observe a hexagonally
    ordered Cs adsorption layer with a mutual adsorbate distance of
    $1.1\,{\rm nm}$. This distance is discussed in terms of a commensurate
    adsorption superstructure which is stabilized by long-range adsorbate
    interactions mediated by Cu(111) surface state electrons. Intermediate
    coverages are characterized by incommensurate superstructures which are
    rotated with respect to the substrate lattice. The rotation angle varies
    with coverage and follows a trend which is consistent with models of
    epitaxial rotation. With increasing coverage the adsorption layers are
    found to rotate toward alignment with the substrate.
  \end{abstract}

  \pacs{06.30.Bp,68.37.Ef,68.55.Jk}

  \maketitle

  \section{Introduction}
    \subsection{Alkali metal adsorption on metal surfaces}
    Investigations of the adsorption of alkali metals on single-crystal metal
    surfaces has a long tradition in surface science. This is partly due to
    potential technological applications like the promotion of catalytic
    reactions, an enhanced oxidation, \cite{jeo_87,mti_90,gfa_98,syd_99} and
    an increase in electron emission rates. \cite{jbt_33,ahs_68} Moreover
    alkali metal atoms are, due to their simple electronic structure,
    candidates for model chemisorption studies as reviewed,
    for instance, in Refs.\ \onlinecite{jpm_78} and \onlinecite{ndl_89}.
    Recently published articles \cite{mba_97,sog_99,agb_01,jpg_04,cco_05} focus on the
    investigation of electronic and dynamic properties of Cs layers on
    Cu(111). Consequently, it is important to study the geometrical structure
    of this adsorbate system. Before addressing the particular adsorbate
    system Cu(111)-Cs we give a general overview of alkali metals adsorbed
    on surfaces. An excellent review on this subject has been given by Diehl
    and McRath. \cite{rdd_96}

    Structural characterization of alkali adsorbate systems has been a central
    issue and resulted in the following general picture of alkali adsorption
    on metal surfaces. Due to charge transfer from the adsorbed alkali atom
    to the substrate, the alkali adatoms become partially charged. \cite{kwa_89}
    The induced dipole moment then causes the alkali atoms to mutually repel
    leading to homogeneous adatom arrangement. As a result, at very low
    coverages low-energy electron diffraction (LEED) patterns
    reveal a ring around the (0,0) spot. \cite{wcf88b,jco_88,dta_91,zli_91,gsl_96}
    The formation of such structures in LEED patterns indicates a superstructure of
    randomly distributed adsorbed atoms (adatoms) with a prevailing mutual
    distance. In contrast, higher coverages lead to sharp diffraction spots
    indicating a periodic superstructure with long-range order.
    At room temperature, superstructures were observed only for commensurate
    phases, {\it e.\,g.}, for Co($10\bar10$)-K,\cite{tma_92} Au(100)-K,\cite{dfs_91}
    and Ni(111)-K.\cite{pka_93} Below room temperature, incommensurate phases
    were reported, for instance, for Ag(111)-K, -Rb, -Cs,\cite{gsl_96} and
    Ni(100)-K. \cite{dfi92a} Rotation of an incommensurate adsorption layer
    (adlayer) with respect to the substrate, with the rotation angle depending
    continuously on the coverage, was observed for various systems, namely
    C(0001)-Cs, \cite{njw_91} Pt(111)-Na, \cite{jco_88} Pt(111)-K, \cite{gpi_88}
    Ru(0001)-Li, \cite{dlg_86} Ru(0001)-Na, \cite{dlg_84} Ag(111)-K, -Rb, -Cs,
    \cite{gsl_96} Cu(100)-K and Ni(100)-K, \cite{dfi92a}, and Rh(100)-Cs. \cite{gbe_87}
    Models which describe the rotational behavior are based on domain wall
    alignment with a symmetry direction of the substrate, \cite{fgr_91,hsh_79,hsh_80}
    higher-order commensurate phases, \cite{dld_85} or response of an elastic
    overlayer to a small-amplitude corrugation of the substrate potential.
    \cite{adn_77,jpt_79}

    Few alkali-substrate combinations lead to island formation at coverages
    below saturation, like Al(111)-Na, -K, -Rb, \cite{jna_93,hbr_95}
    Al(100)-Na, \cite{wbe_95,rny_97} Cu(111)-Na, \cite{nfi_94} Cu(100)-Li,
    \cite{hto_92} Cu(100)-K, \cite{tar_86}, and Ag(100)-K. \cite{smo_90}
    Within the islands, the alkali atoms form commensurate superstructures.
    A density functional theory study \cite{jne_93,cst_94} indicates
    that the adsorbate-substrate interactions exceed the repulsive
    adsorbate-adsorbate interactions at low coverages for these systems.
    Additionally, these systems condense in (higher-order)
    commensurate phases which suggests that the energy gain from forming
    commensurate phases promotes island formation. \cite{rdd_96}

    For substrate surfaces with square or rectangular symmetry the alkali
    atoms occupy adsorption sites which maximize the coordination number to
    the substrate. \cite{san_75,jed_75,cve_89,sam_92,umu_92,smi_93,wbe_95}
    The on-top adsorption site is frequently observed for hexagonally
    close-packed substrate surfaces, for instance, Cu(111)-p$(2\times2)$Cs,
    \cite{sal_83} Al(111)-$(\sqrt3\times\sqrt3)$Rb, \cite{mke_92} and
    Ni(111)-p$(2\times2)$K \cite{dfi92b} to name only a few, and bridge site
    for Rh(111)-p$(2\times2)$Rb. \cite{ssc_96} For the systems with on-top-site
    adsorption, LEED studies showed that the adatoms push their supporting atom
    below the surface layer leading to surface rumpling and an increase of
    the coordination number. \cite{rdd_96}
    \begin{figure*}
      \includegraphics[width=180mm,clip=]{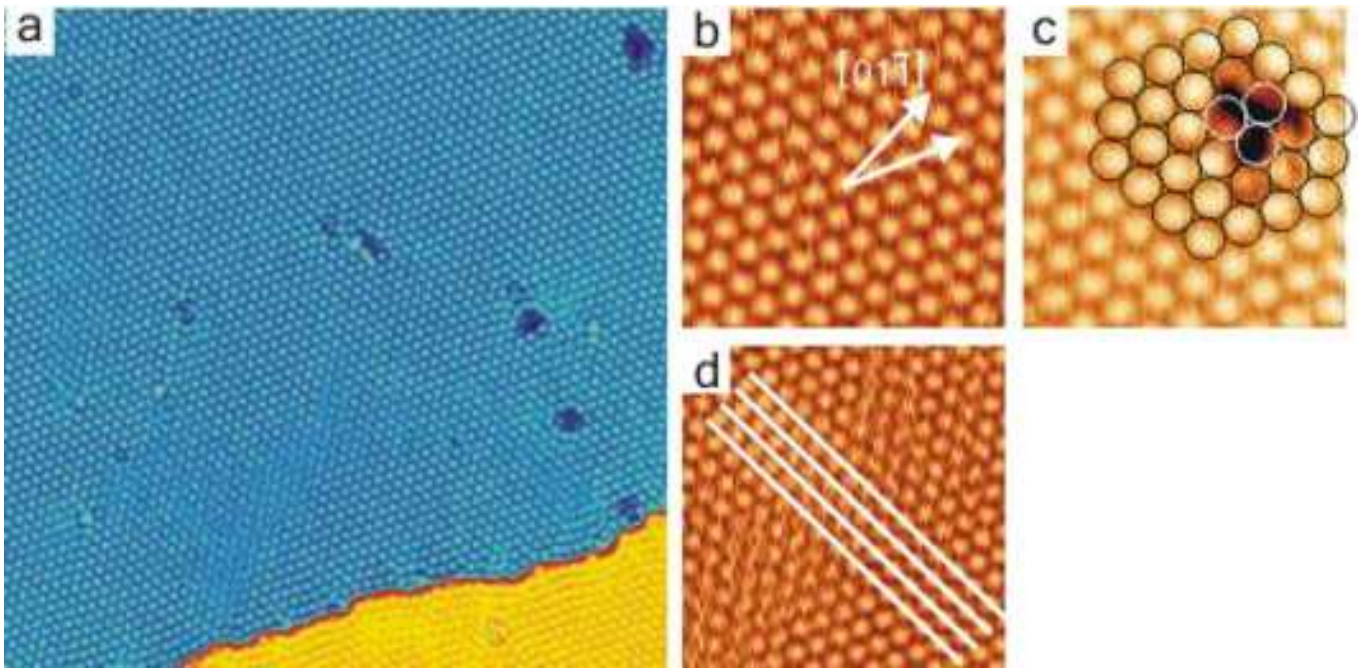}
      \caption[low]{(Color online) a) Constant-current STM image of the
      Cu(111)-Cs surface at 0.05\,ML. The bright circular protrusions are
      assigned to single Cs atoms. ($58\,{\rm nm}\times 58\,{\rm nm}$,
      tunneling parameters: $V=-600\,{\rm mV}$, $I=0.1\,{\rm nA}$; fast scan
      direction is from top to bottom.) b) Close-up view of an area of
      $10\,{\rm nm}\times 10\,{\rm nm}$.
      The indicated $[01\overline{1}]$ direction corresponds to a close
      packing direction of the substrate ($V=-600\,{\rm mV}$, $I=0.1\,{\rm nA}$).
      c) Close-up view ($9\,{\rm nm}\times 9\,{\rm nm}$) contains
      a defect in the top right corner of the image. The black circles
      indicate the position of adjacent Cs atoms, while the white circles
      continue this hexagonal lattice also inside the defect structure. d) Close-up
      view ($14\,{\rm nm}\times 14\,{\rm nm}$) indicating that two adjacent
      Cs domains (upper left and lower right quarter) are mutually shifted
      by half an adatom row.}
      \label{low}
    \end{figure*}

    \subsection{Cs on Cu(111)}
    Two publications report the geometrical structure of Cs adsorbed on
    Cu(111). Lindgren {\it et al.\/} inferred from LEED investigations \cite{sal_83}
    that room-temperature adsorption of Cs on Cu(111) leads to a saturation
    coverage. After completing the first adsorption layer further
    exposure to Cs did not increase the coverage. This adsorption layer
    revealed a p$(2\times 2)$ superstructure, with Cs atoms occupying on-top
    adsorption sites. For lower coverages the ring formation in LEED patterns
    as described above was observed. Further studies of Cu(111)-Cs were
    reported by Fan and coworkers. \cite{wcf88a} They examined adsorption
    structures in the temperature range between $80\,{\rm K}$ and $500\,{\rm K}$.
    As a result the saturation coverage increases to $0.28\,{\rm ML}$ at
    $80\,{\rm K}$, where $1\,{\rm ML}$ is defined as one Cs atom per Cu atom.
    Fan {\it et al.} did not report any commensurate phase for temperatures
    down to $80\,{\rm K}$ except for the already known p$(2\times 2)$
    superstructure. However, they reported orientationally ordered
    incommensurate phases for coverages $\Theta > 0.12\,{\rm ML}$ at $80\,{\rm K}$.
    \begin{figure*}
      \includegraphics[width=180mm,clip=]{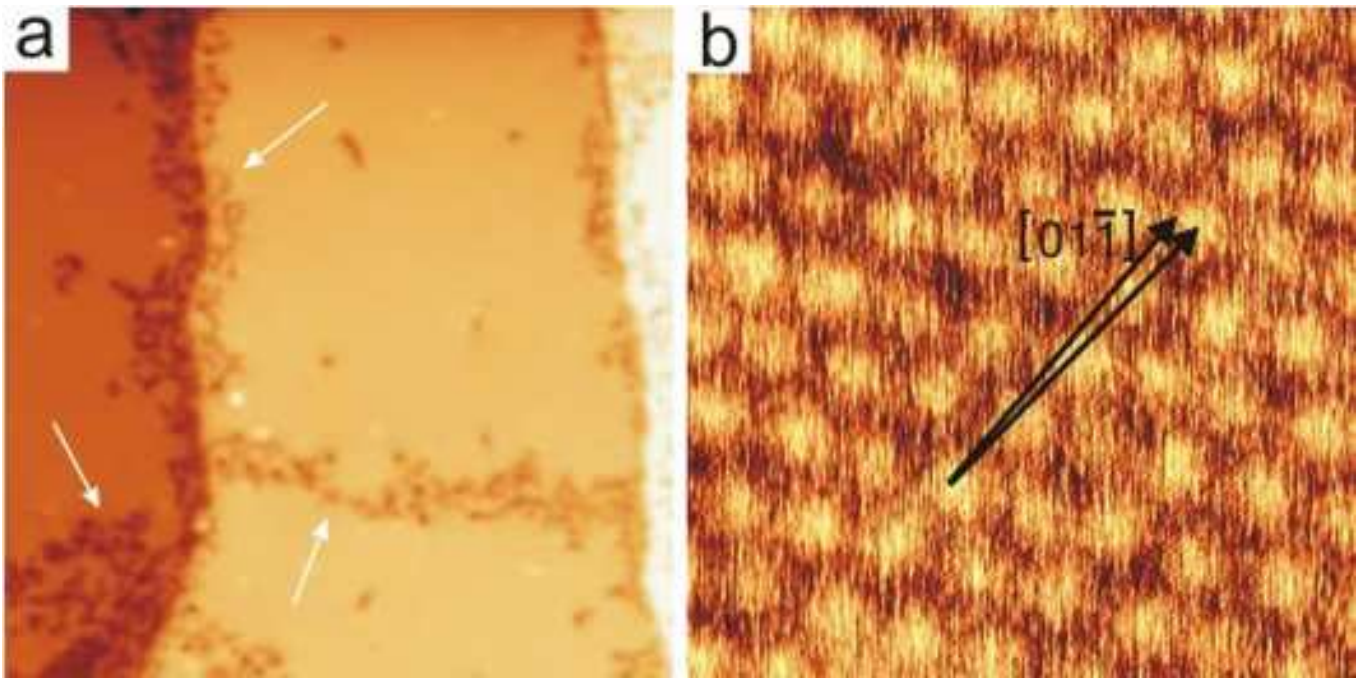}
      \caption[inter]{(Color online) a) Constant-current STM image of
      Cu(111)-Cs at $0.18\,{\rm ML}$. The Cs layer is almost closed, only at
      step edges and occasionally inside the layer, irregularly formed patterns
      (marked by arrows) are observed ($62\,{\rm nm}\times 62\,{\rm nm}$,
      $V=200\,{\rm mV}$, $I=0.1\,{\rm nA}$). b) Close-up view of a scan area
      inside the closed Cs layer displayed in a. The bright circular protrusions
      are assigned to single Cs atoms. Crystallographic orientation of Cu(111)
      is indicated ($4.5\,{\rm nm}\times 4.5\,{\rm nm}$, $V=50\,{\rm mV}$,
      $I=0.1\,{\rm nA}$).}
      \label{inter}
    \end{figure*}

    The cesium-covered Cu(111) surface has attracted much interest recently.
    Excited Cs states in the low-coverage regime have been investigated
    by time-resolved two-photon photoemission experiments to gain insight
    into the bonding and electronic relaxation of alkali atoms on metal
    surfaces. \cite{mba_97,sog_99} The dynamics of quantum well states, which
    are localised in the ultrathin Cs layer on Cu(111), have recently been
    investigated by a combined theoretical and tunneling spectroscopy study.
    \cite{cco_05} Surprisingly, in spite of the substantial interest in their
    dynamic and electronic properties a real-space analysis of the geometric
    structure of Cs films on Cu(111) at various coverages is still lacking.
    In this article we present the results of a scanning tunneling microscopy
    (STM) study of Cu(111)-Cs for different coverages at $9\,{\rm K}$. Atomic
    resolution of the Cs adlayer reveals incommensurate phases, which are
    rotated with respect to the Cu(111) lattice. The rotation angle decreases
    with increasing coverage and vanishes at the p$(2\times 2)$ superstructure.
    At very low coverages a commensurate phase is found, which exhibits mutual
    Cs distances of $1.1\,{\rm nm}$. This distance suggests an electronic
    stabilization of the adsorbate lattice: Adsorbate-induced Friedel oscillations
    of the substrate charge density provided by the Cu(111) surface state
    give rise to an adsorbate-adsorbate distance corresponding to half the
    Fermi wavelength of surface-state electrons.

  \section{Experiment}
  The experiments were performed in ultrahigh vacuum recipients with a base
  pressure of 10$^{-8}$\,Pa. The home-made scanning tunneling microscope was
  operated at room temperature and at $9\,{\rm K}$. Clean Cu(111) surfaces
  were obtained after several cycles of argon ion bombardment and annealing.
  Crystalline order and adsorbate superstructures were monitored with LEED
  and cleanliness was checked with STM. The clean Cu(111) surface was exposed
  to Cs at room temperature. Cesium was evaporated from a commercial dispenser
  \cite{sae_ge} keeping the pressure below $5\times 10^{-8}\,{\rm Pa}$.
  Using a quartz microbalance the deposition rate was monitored to be
  $\approx 0.06\,{\rm ML}\,{\rm min}^{-1}$. We define the coverage as the
  number of Cs atoms per Cu atom, {\it i.\,e.}, one Cs atom per Cu atom
  corresponds to one monolayer (ML). The saturation coverage of Cs at room
  temperature corresponds to the p($2\times 2$) superstructure with a coverage
  of $0.25\,{\rm ML}$. This definition follows Fan {\it et al.} \cite{wcf88a}
  and deviates from the one used by Lindgren and coworkers \cite{sal_83}
  where the saturation coverage is defined as $1\,{\rm ML}$. We find that
  further exposure does not lead to an increase of the surface coverage in
  agreement with the results of Ref.\ \onlinecite{sal_83}. The directions of
  close-packed atoms of the Cu(111) surface were obtained by analyzing LEED
  diffraction patterns with an accuracy of $2^{\circ}$.

  We noticed that during extended LEED measurements of Cu(111)-p$(2\times 2)$Cs the
  superstructure diffraction spots became less intense and less sharp when
  using electron kinetic energies in excess of $100\,{\rm eV}$ on a timescale
  of $1\,{\rm h}$. The dynamic LEED experiments by Lindgren {\it et al.} \cite{sal_83}
  should not be affected by this kind of e-beam damage since the maximum
  electron energy in their experiments was $150\,{\rm eV}$ and the measurements
  were performed within a few minutes.

  \section{Results}
  Typical constant-current STM images of Cu(111)-Cs acquired at room
  temperature (not shown) do not reveal any adsorbate superstructure. We
  inferred the presence of the adsorbate from a noisy tunneling current, which
  we do not usually observe on clean metal surfaces. Further, step edges were
  not imaged as straight lines. Rather, step edges appear frayed in
  constant-current STM images. We attribute these observations to
  mobility and to tip-induced movements of the Cs adatoms. Due to the
  instability of the tunneling junction, tunneling spectroscopy measurements
  were difficult to perform in a reproducible manner at room temperature.
  As a consequence we performed our experiments at $9\,{\rm K}$. The
  data to be presented in the following were acquired at low temperature.

    \subsection{Low coverage: $\Theta=0.05\,{\rm ML}$}
    Figure \ref{low}a shows a representative constant-current STM image of
    Cu(111) covered with $0.05\,{\rm ML}$ Cs revealing an area of more than
    $2500\,{\rm nm}^2$ with a step crossing in the lower right. Hexagonally
    ordered bright circular protrusions cover the whole image. We observed this
    superstructure for many different areas of the sample. The close-up view
    in Fig.\ \ref{low}b shows the hexagonal Cs superlattice
    in more detail. From this image an interatomic distance of $1.1\,{\rm nm}$
    can be determined. The corrugation of the superlattice is
    $\approx 0.03\,{\rm nm}$. The arrows indicate directions of close packing
    of the Cu(111) substrate ($[01\overline{1}]$) and of the adlayer.
    Evidently the Cs layer is rotated by $\approx 23^{\circ}$ with respect to
    the Cu(111) surface. On the right side of Fig.\ \ref{low}a defects of
    the adsorption layer are observed. We attribute these defects to
    imperfections of the Cs layer, {\it i.\,e.}, to missing Cs adatoms. To
    corroborate this assumption a close-up view of the defect in the upper
    right corner of Fig.\ \ref{low}a is shown in Fig.\ \ref{low}c. The black
    circles indicate the positions of Cs atoms in the layer, while the white
    circles continue the hexagonal lattice inside the defect structure. From
    this image we infer that the dark area corresponds to missing Cs atoms.

    The hexagonal order of the adatoms extends over large areas. In some
    regions, atomic resolution of the adsorbate layer is lost during scanning.
    These regions appear as blurred stripes in the fast scanning direction
    (from top to bottom in Fig.\ \ref{low}d). In $\approx 40\,\%$
    of the cases where the blurred stripes appear, adjacent Cs-covered areas
    are shifted with respect to each other by half a superlattice constant
    as indicated by white lines in Fig.\ \ref{low}d. We attribute this
    observation to adjacent Cs adsorption domains. In both domains Cs atoms
    reside at stable adsorption sites, while in the region between the domains
    no such stable adsorption site is available. As a consequence, Cs atoms
    in these regions are prone to be moved by the tip leading to the observed
    loss of atomic resolution.

    \subsection{Intermediate coverage: $\Theta=0.15-0.20\,{\rm ML}$}
    Higher Cs coverages led to an increase of the radius of the ring structure
    in the LEED pattern at room temperature pointing to a decrease of the
    average separation between Cs atoms. \cite{sal_83} A typical
    constant-current STM image of Cu(111) covered with $0.18\,{\rm ML}$ Cs
    is shown in Fig.\ \ref{inter}a. We observe an almost closed Cs layer, which
    is disrupted by small and irregularly shaped indentations. Occasionally,
    these structures occur within a closed Cs layer, but most frequently they
    are observed at step edges. The apparent depth of the indentations depends
    on the applied voltage. Atomic resolution of flat areas of the Cs layer
    is presented in Fig.\ \ref{inter}b. Again, we identify the bright
    protrusions as Cs atoms.
    Distances between nearest neighbors are $0.60\,{\rm nm}$ for this
    coverage. The corrugation of the superlattice is $0.003\,{\rm nm}$ which
    is considerably smaller than for $\Theta=0.05\,{\rm ML}$. We attribute this
    observation to the Cs adatoms being more densely packed at higher coverage.
    Comparing with the Cu(111) substrate lattice we find a rotation angle of
    the adsorbate layer of $4^{\circ}$. We obtained the same values for the
    interatomic distance, the rotation angle, and the corrugation at various
    areas on the sample.

    For Cs coverages of $0.15\,{\rm ML}$ and $0.20\,{\rm ML}$ Cs-Cs
    distances of $0.66\,{\rm nm}$ and $0.57\,{\rm nm}$ occur, respectively.
    The adsorbate layers are rotated with respect to the substrate lattice by
    $17^{\circ}$ and
    \begin{figure}
      \includegraphics[width=85mm,clip=]{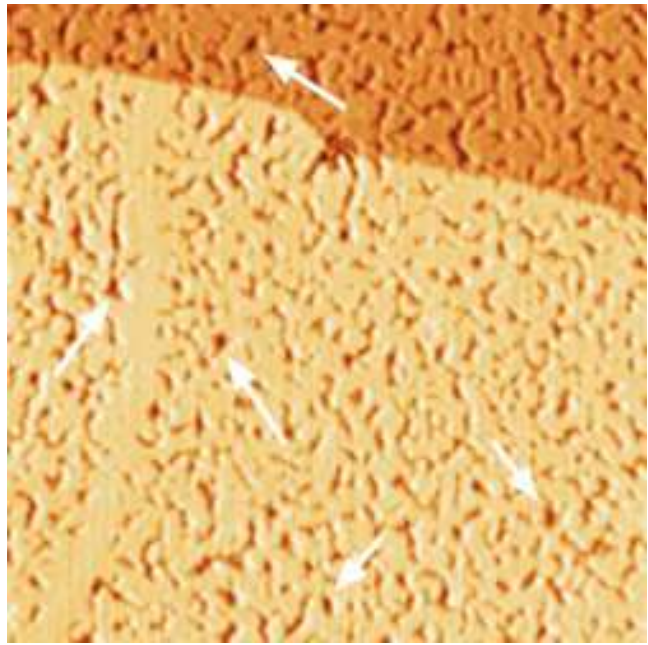}
      \caption[p22]{(Color online) Constant-current STM image of
      Cu(111)-p($2\times 2$)Cs at $9\,{\rm K}$
      ($124\,{\rm nm}\times 124\,{\rm nm}$, $V=250\,{\rm mV}$, $I=0.2\,{\rm nA}$).}
      \label{p22}
    \end{figure}
    $0^{\circ}$, respectively. Upon increasing the coverage the Cs adlayer becomes more and
    more disrupted. Instead of a closed adsorption layer numerous small Cs
    islands are observed (see the results for saturation coverage).

    \subsection{Saturation coverage: $\Theta=0.25\,{\rm ML}$}
    At $0.25\,{\rm ML}$ the room temperature LEED pattern reveals a clear
    $(2\times 2)$ superstructure. A typical constant-current STM image of
    this surface at $9\,{\rm K}$ is displayed in Fig.\ \ref{p22}. The
    adsorption layer is characterized by a Cs film revealing an increased
    number of imperfections compared to lower coverages.
    These imperfections, which appear as dark indentations at the given
    tunneling voltage exhibit a variety of sizes and shapes. We found that
    the dimensions of the imperfections (apparent height and lateral size)
    depend on the applied tunneling voltage.

  \section{Discussion}
    \subsection{Low coverage: $\Theta=0.05\,{\rm ML}$}
    The measured interatomic distance of $1.1\,{\rm nm}$ and the rotation angle
    of $23^{\circ}$ match well a $(\sqrt{19}\times\sqrt{19})\,R\,23.4^{\circ}$
    commensurate phase. Stabilization of this commensurate superstructure may
    arise from long-range adsorbate-adsorbate interactions mediated by
    substrate electrons. Lau and Kohn \cite{khl_77} predicted that adsorbates
    may interact via Friedel oscillations \cite{jfr_58} through the fact that
    the binding energy of one adsorbate depends on the substrate electron
    density, which oscillates around the other adsorbate. Lau and Kohn later
    found for a two-dimensional electron gas that these interactions depend
    on distance, $r$, as $r^{-2}\,\cos(2\,k_{\rm F}\,r)$ where $k_{\rm F}$ is
    the Fermi vector. \cite{khl_78} The Cu(111) surface hosts an electronic
    surface state which is a model system for a two-dimensional free
    electron gas. The Fermi vector of this surface state is
    $k_{\rm F}\approx 2.2\,{\rm nm}^{-1}$ giving rise to Friedel
    oscillations with the Fermi wavelength
    $\lambda_{\rm F}=2\pi k_{\rm F}^{-1}\approx 2.9\,{\rm nm}$.\cite{phy_00} Indications of
    such a long-range interaction between strongly bonded sulfur atoms on a
    Cu(111) surface have been first reported in Ref.\ \onlinecite{ewa_98}.
    The first quantitative study of a long-range interaction mediated
    by a two-dimensional nearly free electron gas was reported by Repp
    {\it et al.} \cite{jre_00} for Cu(111)-Cu and later for Cu(111)-Cu,
    Cu(111)-Co, and Ag(111)-Co by Knorr {\it et al.}. \cite{nkn_02}
    The closest separation between
    two Cu adatoms was $1.25\,{\rm nm}$ corresponding roughly to
    $\lambda_{\rm F}/2$ of the Cu(111) surface state. An atomic
    superlattice was also observed for adsorbed Ce atoms on a Ag(111) surface.
    \cite{fsi04a} The observed $3.2\,{\rm nm}$ periodicity of the superlattice
    was assigned to the interaction of surface-state electrons with the Ce
    adatoms.

    The interaction energy between adsorbates as mediated by surface state
    electrons was found to be \cite{phy_00}
    \begin{equation}
      \label{energy}
      \Delta E_{\rm int}(r)\simeq -E_{\rm F}\,\left(\frac{2\sin\delta_{\rm F}}{\pi}\right)^2\,\frac{\sin(2k_{\rm F}r+2\delta_{\rm F})}{(k_{\rm F}r)^2}
    \end{equation}
    where $E_{\rm F}$ denotes the Fermi energy measured from the bottom of
    the surface-state band and $\delta_{\rm F}$ is the Fermi-level phase
    shift, which depends on the scatterer. \cite{gho_94} Inserting the
    experimentally observed mutual Cs distance of $(1.1\pm 0.1)\,{\rm nm}$
    into Eq.\ (\ref{energy}) leads to a phase shift of
    $\delta_{\rm F}=(0.43\pm 0.08)\pi$. This is close to the value of
    $\delta_{\rm F}=(0.50\pm 0.07)\pi$ for Cu and Co on Cu(111). \cite{nkn_02}
    The similarity of these values indicates that the phase shift does not
    vary appreciably among these scatterers with different chemical nature.
    Since the value of $\delta_{\rm F}$ in our case, {\it i.\,e.}, Cu(111)-Cs,
    is similar to the phase shifts obtained from the systems above, we have
    additional evidence that the interaction between Cs adatoms at low
    coverages is mediated by the Cu(111) surface state.

    \subsection{Intermediate coverage: $\Theta=0.15-0.20\,{\rm ML}$}
\begin{figure}
      \includegraphics[width=85mm,clip=]{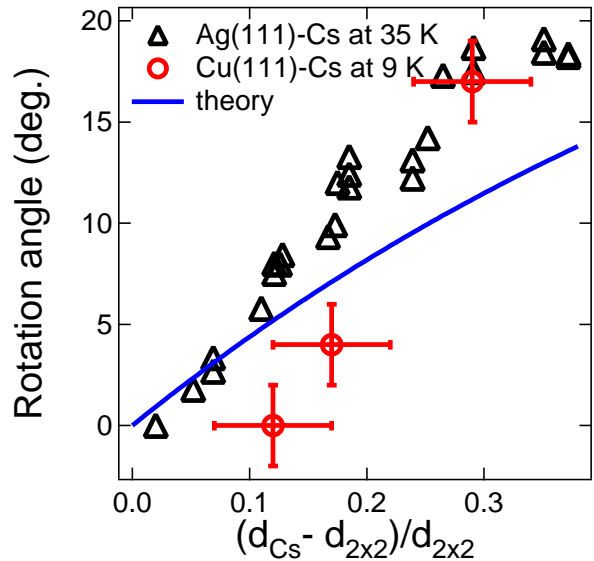}
      \caption[rot]{(Color online) Rotation angle of a Cs adlayer on Cu(111)
      and Ag(111) versus misfit. Circles: Rotation angle of Cs adlayer on
      Cu(111), triangles: Rotation angle of Cs adlayer on Ag(111) as adapted
      from Ref.\ \onlinecite{gsl_96}, solid line: Model calculation as
      described in the text.}
      \label{rot}
    \end{figure}

    For the Cs superlattices observed at intermediate coverages, we do not
    find a coincidence of the substrate and adsorbate lattice. Consequently,
    on the basis of our STM investigation, we propose the adsorption layers
    in the intermediate coverage regime to be incommensurate. We further
    observed that the incommensurate adsorption phases are rotated with
    respect to the substrate lattice. The rotation angle of the incommensurate
    Cs adsorbate layers on Cu(111) as a function of the misfit
    $(d_{\rm Cs}-d_{2\times 2})/d_{2\times 2}$ is shown in Fig.\ \ref{rot}
    (circles). Here $d_{\rm Cs}$ denotes the nearest-neighbor distance of the
    Cs adsorption layer and $d_{2\times 2}$ the distance of the Cs atoms in
    the p($2\times 2)$ superstructure. The lattice constant of the Cu(111)
    substrate $d_{\rm Cu}\approx 0.255\,{\rm nm}$ leads to a
    nearest-neighbor distance of Cs adatoms in the p($2\times 2$)
    superstructure of $d_{2\times 2}\approx 0.511\,{\rm nm}$.
    We added data adapted from Ref.\ \onlinecite{gsl_96} displaying the
    rotation angles for incommensurate phases of Ag(111)-Cs at 35\,K
    obtained by LEED (triangles). While a general trend of increasing
    rotation angle with increasing misfit (decreasing coverage) is observed
    for the two adsorbate systems, the rotation of adsorbate layers starts
    at larger misfits in the case of Cu(111)-Cs.

    The rotation of adsorbed overlayers relative to the substrate has been
    observed previously for other adsorption systems. \cite{fgr_92,fgr_91}
    As a consequence, several explanations of the rotational behavior of
    adlayers have been put forward. \cite{gsl_96}
    A simple geometrical model first applied by Doering \cite{dld_85} proposes
    that the overlayer forms a higher-order commensurate phase, where
    the superstructure unit cell is much larger than the unit cell of the
    substrate surface. This leads to a large number of possible rotation
    angles for each misfit. It remains unclear why a specific orientation of
    the higher-order commensurate phase arises.
    Another geometrical model by Bohr and Grey \cite{fgr_91} makes use of
    domain formation. The domains in question are the result of a Moir\'{e}
    pattern: Since the adlayer has larger interatomic spacings than the
    substrate there are domains where the adatoms are nearly in phase with
    substrate atoms; the separating areas, where the displacement is larger,
    are called domain walls. \cite{fgr_92}
    The assumption of the model is that the layer aligns to the substrate in
    such a way that the domain walls are oriented in a high-symmetry direction
    of the substrate or of the layer to minimize energy.
    Within the model by Novaco and McTague \cite{adn_77,jpt_79} the rotation
    angle of the adsorbate layer with respect to the substrate lattice is
    determined on the basis of a weak adsorbate-substrate interaction. Within
    a semiclassical approximation Novaco and McTague found the rotation angle
    to depend on the ratio of the longitudinal and transverse speed of sound
    in the adsorbate layer, $c_{\rm L}$ and $c_{\rm T}$, respectively, and
    upon the ratio of the lattice constants of the substrate, $d_{2\times2}$,
    and the adsorbate layer, $d_{\rm Cs}$:
    \begin{equation}
      \cos\theta = \frac{1 + z^2 (1 + 2\eta)}{z\left[2 + \eta (1 + z^2)\right]}
      \label{mac}
    \end{equation}
    with $\eta = (c_{\rm L}/c_{\rm T})^2 - 1$ and $z = d_{2\times 2}/d_{\rm Cs}$.
    While the lattice constants are known, little is known about the sound
    velocities in thin films. For alkali metals on noble metal surfaces, phonon
    dispersion curves were measured for Na, K and Cs on Cu(001).
    \cite{gbe_92,ehu_96,gwi_00} From the phonon dispersion relations of Cs
    on Cu(001), \cite{gwi_00} we know that $c_{\rm L}/c_{\rm T}=1.2$, which
    leads to $\cos\theta>1$ according to Eq.\ (\ref{mac}). We therefore
    calculated the ratio of sound velocities of the Cs adsorption layer by
    solving the two-dimensional dynamical equation
    \begin{equation}
      m\,\frac{{\rm d}^2 x_{n\alpha i}}{{\rm d}t^2} + \sum_k\,\Phi_{n\alpha i}^{m\beta k}\,x_{m\beta k} = 0 \quad (i = 1,2),
    \end{equation}
    where $x_{n\alpha i}$ denotes the excursion of atom $\alpha$ in unit cell
    $n$ from equilibrium in direction $i$, $m$ is the adatom mass, and
    \begin{equation}
      \Phi_{n\alpha i}^{m\beta k} = \frac{\partial^2 \Phi}{\partial x_{n\alpha i}\,\partial x_{m\beta k}}
    \end{equation}
    are the coupling constants, which are calculated as the second derivative
    of the interaction potential. For the latter we assume a dipole potential.
    Numerical calculations showed that including interactions ranging further
    than the nearest neighbor do not change the result significantly. Since
    the rotation angle depends only upon the ratio of the sound velocities
    the actual dipole moment of Cs atoms on Cu(111) is not required for the
    calculation. We find $c_{\rm L}/c_{\rm T}=\sqrt {11}$ and plot $\theta$
    according to Eq.\ (\ref{mac}) in Fig.\ \ref{rot} as a full line. The
    model reproduces the experimentally observed trend correctly (the larger
    the misfit, the larger the angle of rotation). However, the experimental
    result that rotation starts not until a certain misfit is exceeded is
    not accounted for by the model. This shortcoming of the model close to
    commensurate phases is known. A similar effect was reported for K and Rb
    adsorbed on Ag(111). \cite{gsl_96} The authors interpreted their observation
    as follows: Near commensurate phases where the misfit is small the domains
    become larger and more adatoms are nearly in phase with substrate atoms.
    This leads to a stronger interaction of substrate and adlayer, locking
    the adlayer to the substrate orientation. A theoretical model developed
    by Shiba \cite{hsh_79,hsh_80} takes this into account and predicts that
    for a small misfit the domain walls of the layer remain aligned to a
    symmetry direction of the layer up to a critical misfit. For larger
    misfits, the domain walls become too weak to keep the alignment, and the
    layer is rotated versus the substrate.

    \subsection{Saturation coverage: $\Theta=0.25\,{\rm ML}$}
    The peculiar structure of the Cs adsorbate film as seen in constant-current
    STM images of Cu(111)-p($2\times 2$)Cs can be explained in terms of
    island growth. As mentioned in the introduction
    alkali metal adsorption on metal surfaces goes hand in hand with the
    creation of dipole moments due to charge transfer
    processes. At low coverages the dipole-dipole interaction leads to a
    repulsive Cs-Cs interaction. With increasing coverage, the Cs adatoms
    depolarize and the repulsion decreases. \cite{kwa_89} Reaching a
    sufficiently high coverage the interatomic distance is small enough to
    favor metallic bonds \cite{sal_80} and small islands are formed. With
    increasing size the islands eventually come close to each other and
    coalesce, thereby leaving behind unoccupied substrate areas.

    An additional contribution to the formation of the observed Cs film
    structure may be due to different thermal expansion coefficients. Copper
    contracts by $\approx 0.3\,\%$
    when cooled from room temperature to $4\,{\rm K}$, \cite{gkw_68} {\it i.\,e.},
    the lattice constant reduces to $2.54\,{\rm\AA}$ at $4\,{\rm K}$. For
    (bulk) Cs published thermal expansion coefficients have not been found.
    Taking corresponding data for potassium \cite{drs_74} as an approximation
    for the thermal expansion coefficient of Cs we estimate that the Cs film
    contracts by $\approx 2\,\%$
    upon cooling from room temperature to $4\,{\rm K}$. Consequently, the Cs
    lattice constant reduces to $\approx 5\,{\rm\AA}$. Supposing that at
    room temperature the Cs layer covers the complete surface, which is corroborated
    by our room temperature STM and LEED studies, the different contraction
    of the Cs adsorption layer and the Cu surface leads to a coverage of $96\,\%$
    of the surface at $4\,{\rm K}$, leaving $4\,\%$
    of the surface uncovered. Analysing a variety of constant-current STM
    images of the p$(2\times 2)$Cs superstructure we find that $\approx 82\,\%$
    of the surface was covered. As a consequence, the different thermal
    contraction of the Cs adsorption layer and the Cu surface may contribute
    to the observed layer structure although cannot explain the entire
    effect.

  \section{Summary}
  We used low-temperature STM to investigate Cu(111)-Cs at various Cs coverages.
  For very low coverages the Cs adsorption layer reveals hexagonal order with
  a nearest-neighbor distance of $1.1\,{\rm nm}$. We interpret this adsorption
  phase as a commensurate superstructure stabilized by long-range interactions
  mediated by surface state electrons. For higher coverages
  we find incommensurate Cs superlattices, which are rotated with respect to
  the Cu(111) substrate lattice. In contrast to the high degree of order
  at low and intermediate coverages, the p$(2\times 2)$Cs superstructure at
  saturation coverage exhibits a fairly large defect density.

  \noindent
  Financial support by the Deutsche Forschungsgemeinschaft via the
  Schwerpunktprogramm 1093 is gratefully acknowledged.

\end{document}